\documentclass[12pt,a4wide,epsf]{article}
\usepackage{epsf}

\usepackage{graphicx}

\newcommand{\insertplot}[5]{\begin{figure}
 \hfill\hbox to 0.05in{\vbox to #5in{\vfill
 \inputplot{#1}{#4}{#5}}\hfill}
 \hfill\vspace{-.1in}
 \caption{#2}\label{#3}
 \end{figure}}
 \newcommand{\inputplot}[3]{
 \special{ps: plotfile #1}
}
\newcounter{fig}   

\newcommand{\vphi}{\varphi}

\newcommand{\sqdetg}{\sqrt{-g}}

\begin{document}

\title{
Five-Dimensional Charged Rotating Black Holes}
 \vspace{1.5truecm}
\author{
{\bf Jutta Kunz}\\
Institut f\"ur Physik, Universit\"at Oldenburg, Postfach 2503\\
D-26111 Oldenburg, Germany\\
{\bf Francisco Navarro-L\'erida}\\
Dept.~de F\'{\i}sica At\'omica, Molecular y Nuclear, Ciencias F\'{\i}sicas\\
Universidad Complutense de Madrid, E-28040 Madrid, Spain\\
{\bf Anna Katinka Petersen}\\
Institut f\"ur Physik, Universit\"at Oldenburg, Postfach 2503\\
D-26111 Oldenburg, Germany
}

\vspace{1.5truecm}

\date{\today}

\maketitle
\vspace{1.0truecm}

\begin{abstract}
We consider charged rotating black holes 
in 5-dimensional Einstein-Maxwell theory.
These black holes are asymptotically flat,
they possess a regular horizon of spherical topology
and two independent angular momenta
associated with two distinct planes of rotation.
We discuss their global and horizon properties,
and derive a generalized Smarr formula. 
We construct these black holes numerically, 
focussing on black holes with a single angular momentum,
and with two equal-magnitude angular momenta.
\end{abstract}

\vfill\eject

\section{Introduction}

In Einstein-Maxwell (EM) theory in $3+1$ dimensions
the unique family of stationary asymptotically flat black holes
comprises the rotating Kerr-Newman (KN) and Kerr black holes
and the static Reissner-Nordstr\"om (RN) and Schwarzschild black holes.
EM black holes are uniquely determined
by their mass, their electric and magnetic charge,
and their angular momentum
\cite{unique}.

The generalization of black hole solutions to higher
dimensions was pioneered by Tangherlini
\cite{tangher} for static vacuum black holes,
and by Myers and Perry \cite{MP} for stationary vacuum black holes.
Whereas Myers and Perry \cite{MP} also obtained
charged static black holes in higher dimensional EM theory,
higher dimensional charged rotating black holes have not yet
been obtained in pure EM theory, 
although such black holes are known
for some low energy effective actions related to string theory
\cite{string1,string2}.
Higher dimensional black holes received much interest in recent years also
with the advent of brane-world theories, raising the possibility
of direct observation in future high energy colliders
\cite{exp}.

The uniqueness theorems of EM theory in $3+1$ dimensions
cannot be generalized to higher dimensions, unless one makes restrictions,
in particular with respect to the horizon topology 
\cite{ida}.
Indeed, there exist black rings, i.e., solutions with horizon
topology of a torus \cite{blackrings}.
Restricting to black holes with horizon topology of a sphere,
we here consider charged rotating EM black holes in $4+1$ dimensions,
which are asymptotically flat, 
possess a regular event horizon, 
and in general possess two independent angular momenta,
associated with two distinct planes of rotation.
Recently, such black holes have been studied perturbatively, 
in the limit of small charge \cite{frolov}.

We investigate the physical properties of these black holes.
We obtain their mass $M$, their angular momenta $J_1$ and $J_2$,
and their electric charge $Q$ from an asymptotic expansion. 
For generic values of the charge, 
their gyromagnetic ratios deviate from the perturbative values,
$g_i=3$ \cite{frolov}.
From an expansion at the horizon
we determine their horizon area $A_{\rm H}$, their surface gravity $\kappa$,
their horizon velocities $\Omega_1$ and $\Omega_2$, 
and their horizon electrostatic potential $\Phi_{\rm H}$.
We then show, that these black holes satisfy
a generalized Smarr formula \cite{smarr}
\begin{equation}
\frac{2}{3}M=
\frac{\kappa A_{\rm H}}{8\pi G}+\Omega_1 J_1 + \Omega_2 J_2
+ \frac{2}{3}\Phi_{\rm H}\, Q
\ . \label{mass} \end{equation}

To construct these charged rotating black holes numerically,
for finite values of the charge,
we employ an Ansatz for the metric which is inspired by
the parametrization of Myers and Perry \cite{MP,frolov},
but based on bi-azimuthal isotropic coordinates. 
This parametrization of the metric
corresponds to a generalization of the Ansatz employed
previously for rotating black holes in $3+1$ dimensions,
well suited for numerical work \cite{kkrot,kkrot_c}.
The Ansatz for the gauge potential
follows from the symmetries \cite{frolov}.

In section 2 we recall the EM action, 
and present the stationary axially symmetric Ansatz.
We discuss the black hole properties in section 3,
and derive the generalized Smarr formula in section 4.
We present some numerical results in section 5,
focussing on black holes with a single angular momentum
and with two equal-magnitude angular momenta.

\section{Action and Ansatz}

We consider the 5-dimensional Einstein-Maxwell action
\begin{equation}
S=\int \left ( \frac{R}{16\pi G}
  - \frac{1}{4} F_{\mu\nu} F^{\mu\nu}
 \right ) \sqrt{-g} d^5x
\ , \label{action} 
\end{equation}
with curvature scalar $R$,
Newton's constant $G$,
and field strength tensor
$
F_{\mu \nu} = 
\partial_\mu A_\nu -\partial_\nu A_\mu $,
where $A_\mu $ denotes the gauge potential.

Variation of the action with respect to the metric and the gauge potential
leads to the Einstein equations and the gauge field equations,
respectively,
\begin{equation}
G_{\mu\nu}= R_{\mu\nu}-\frac{1}{2}g_{\mu\nu}R = 8\pi G T_{\mu\nu}
\ , \label{ee} 
\end{equation}
\begin{equation}
\nabla_\mu F^{\mu\nu}  = 0 \ .
\label{feqA} 
\end{equation}

To obtain stationary black hole solutions,
representing charged generalizations of the 5-dimensional 
Myers-Perry solutions \cite{MP},
we consider black hole space-times with bi-azimuthal symmetry,
implying the existence of three commuting Killing vectors,
$\xi = \partial_t$, $\eta_1=\partial_\varphi$, and $\eta_2=\partial_\psi$
\cite{MP,frolov}.
We employ a parametrization for the metric,
based on bi-azimuthal isotropic coordinates, 
well suited for numerical work,
corresponding to a generalization of the Ansatz employed 
previously for rotating black holes in $3+1$ dimensions \cite{kkrot,kkrot_c}
\begin{eqnarray}
ds^2 &=& -fdt^2+\frac{m}{f}\left( dr^2+r^2 d\theta^2 \right)
 +\frac{p}{f}\, r^6 \sin^2\theta \cos^2\theta 
 \left( \omega_2 d\vphi - \omega_1 d\psi \right)^2 \nonumber \\
      && +\frac{l}{f}\, r^2 \sin^2\theta
          \left(d\vphi-\frac{\omega_1}{r}dt\right)^2
          +\frac{n}{f}\, r^2 \cos^2\theta
          \left(d\psi-\frac{\omega_2}{r}dt\right)^2 
          \ . \label{metric} \end{eqnarray}
The metric has two orthogonal 2-planes of rotation, 
corresponding to $\theta=0$ and $\theta=\pi/2$,
where elementary flatness requires
\begin{equation}
\left. \left( {m-p\, r^4 \omega_2^2} \right) \right|_{\theta=0}=
\left. {l} \right|_{\theta=0}
\ , \ \ \
\left. \left( {m-p\, r^4 \omega_1^2} \right) \right|_{\theta=\pi/2}=
\left. {n} \right|_{\theta=\pi/2}
\ . \end{equation}
The gauge potential is parametrized by \cite{frolov}
\begin{equation}
A_{\mu}dx^{\mu}=A_t dt+A_{\vphi}d\vphi+A_{\psi}d\psi \ . 
\end{equation}
All metric and gauge field functions depend on $r$ and $\theta$ only.

\section{Black hole properties}

The mass $M$ and the angular momenta $J_i$ of the black hole \cite{MP,string3},
are obtained from the asymptotic expansion for the metric \cite{long},
\begin{equation}
f=1-\frac{8\, G M }{3\pi r^2} + O\left(\frac{1}{r^4}\right) \ , \ \ \
\omega_i = \frac{4\, G J_i }{\pi r^3} + O\left(\frac{1}{r^5}\right) \ , \ \ \
\end{equation}
while the charge $Q$ and the magnetic moments ${\cal M}_i$ 
\cite{frolov,string3},
are obtained from the asymptotic expansion for the gauge potential \cite{long},
\begin{eqnarray}
A_t=\frac{Q}{4 \pi^2 r^2} + O\left(\frac{1}{r^4}\right) \ , \ \ \ &&
A_\vphi=-\frac{{\cal M}_1 \sin^2\theta}{4 \pi^2 r^2} 
 + O\left(\frac{1}{r^4}\right) \ , \ \ \ \nonumber\\ &&
A_\psi=-\frac{{\cal M}_2 \cos^2\theta}{4 \pi^2 r^2} 
 + O\left(\frac{1}{r^4}\right) 
\ , \end{eqnarray}
and yield the gyromagnetic ratios $g_i$ \cite{frolov}
\begin{equation}
{\cal M}_1=g_1 \frac{Q J_1}{2M} \ , \ \ \
{\cal M}_2=g_2 \frac{Q J_2}{2M} 
\ . \end{equation}

The event horizon resides at a surface of constant radial coordinate,
$r=r_{\rm H}$ \cite{MP,frolov}.
The Killing vector
\begin{equation}
\chi=\xi + \Omega_1 \eta_1 + \Omega_2 \eta_2 
\label{Killing} \end{equation}
is orthogonal to and null on the horizon,
where $\Omega_1$ and $\Omega_2$ denote the horizon angular velocities
with respect to rotation in the $\theta=0$ and $\theta=\pi/2$ plane,
respectively \cite{MP,frolov},
\begin{equation}
\Omega_1= 
 \left. \frac{g_{t \psi} g_{\vphi \psi} - g_{t \vphi} g_{\psi \psi}}
               {g_{\vphi \vphi} g_{\psi \psi} - g_{\vphi \psi}^2}
 \right|_{r_{\rm H}}
= \frac{\omega_1(r_{\rm H})}{r_{\rm H}}
\ , \ \ \
\Omega_2= 
 \left. \frac{g_{t \vphi} g_{\vphi \psi} - g_{t \psi} g_{\vphi \vphi}}
               {g_{\vphi \vphi} g_{\psi \psi} - g_{\vphi \psi}^2}
 \right|_{r_{\rm H}}
= \frac{\omega_2(r_{\rm H})}{r_{\rm H}}
\ . \label{Omega} \end{equation}
The area ${A}_{\rm H}$ of the black hole horizon is given by
\begin{equation}
{A}_{\rm H} = (2 \pi)^2 r_{\rm H}^3
\int_0^{\pi/2}  d\theta \sin \theta  \cos \theta
\left. \sqrt{\frac{m}{f^3} \left[ ln + p r^4  
\left(l \sin^2 \theta \, \omega_1^2 + n \cos^2 \theta \omega_2^2 \right)
\right] } \, \right|_{r_{\rm H}}
\ , \label{area} \end{equation}
defining the area parameter $r_\Delta$ via
${A}_{\rm H}= 2 \pi^2 r_\Delta^3$.
As required by the zeroth law of black hole mechanics,
the surface gravity $\kappa$ \cite{wald}
\begin{equation}
\kappa^2 = - \frac{1}{2} \left. 
 (\nabla_\mu \chi_\nu)(\nabla^\mu \chi^\nu) \right|_{r_{\rm H}}
= 
 \left. \frac{f^2}{ \left(r-r_{\rm H}\right)^2 m}
 \right|_{r_{\rm H}}
\ , \label{sgwald} \end{equation}
is constant at the horizon. The expansion at the horizon \cite{long}
shows furthermore, that the electrostatic potential $\Phi$
is constant at the horizon as well,
\begin{equation}
\Phi(r_{\rm H}) = \left. \chi^\mu A_\mu \right|_{r_{\rm H}}= 
\left. \left( A_t + \Omega_1 A_\vphi + \Omega_2 A_\psi
\right) \right|_{r_{\rm H}} = 
\Phi_{\rm H}
\ . \label{esp} \end{equation}

\section{Mass formula}

To obtain the general mass formula, 
we recall the Komar expressions for the mass $M$
and the angular momenta $J_i$ 
\cite{wald}
\begin{equation}
M = - \frac{1}{16 \pi G}\frac{3}{2}\int_{S_{\infty}^{3}} \alpha \ , \ \ \
J_i =  \frac{1}{16 \pi G}\int_{S_{\infty}^{3}} \beta_{(i)} \ , 
\label{MJ} \end{equation}
where
\begin{equation}
\alpha_{\mu \nu \rho} =\epsilon_{\mu \nu \rho \sigma \tau}
\nabla^{\sigma} \xi^{\tau} \ , \ \ \
\beta_{(i)\mu \nu \rho} =\epsilon_{\mu \nu \rho \sigma \tau}
\nabla^{\sigma} \eta_i^{\tau}
\ . \end{equation}
The horizon mass $M_{\rm H}$ and horizon angular momenta $J_{i,\rm H}$
are obtained by performing the integrations in Eqs.~(\ref{MJ}) at the horizon,
i.e., by replacing $S_{\infty}^{3}$ by $S_{\rm H}^{3}$, 
where we assume the horizon topology of a sphere.
Noting that
\begin{eqnarray}
 M= 
 M_{\rm H} - \frac{1}{16 \pi G} \frac{3}{2} \int_{\Sigma} d\alpha 
= M_{\rm H}-\frac{1}{8 \pi G} \frac{3}{2} \int_{\Sigma}R^{\, t}_{\ t}
\sqdetg dr d\theta d\vphi d\psi
\ , \end{eqnarray}
\begin{eqnarray}
J_1= J_{1,\rm H} + \frac{1}{16 \pi G}\int_{\Sigma} d\beta_1 
= J_{1,\rm H}+\frac{1}{8 \pi G} \int_{\Sigma}R^{\, t}_{\ \vphi}
\sqdetg dr d\theta d\vphi d\psi
\ , \end{eqnarray}
\begin{eqnarray}
J_2= J_{2,\rm H} + \frac{1}{16 \pi G}\int_{\Sigma} d\beta_2 
= J_{2,\rm H}+\frac{1}{8 \pi G} \int_{\Sigma}R^{\, t}_{\ \psi}
\sqdetg dr d\theta d\vphi d\psi
\ , \end{eqnarray}
and taking into account the Killing vector 
$\chi=\xi + \Omega_1 \eta_1 + \Omega_2 \eta_2 $ 
to reexpress the horizon mass \cite{wald}, 
\begin{equation}
\frac{2}{3} M_{\rm H}=   \frac{\kappa A_{\rm H}}{8 \pi G} 
+ \Omega_1 J_{1,\rm H} + \Omega_2 J_{2,\rm H}
\ , \end{equation}
we define the integral $I$,
\begin{eqnarray}
 I &=& \frac{2}{3} M -  \frac{\kappa A_{\rm H}}{8 \pi G} 
 - \Omega_1 J_{1,\rm H} - \Omega_2 J_{2,\rm H}  \nonumber \\
 &=& -\frac{1}{8 \pi G}\int_{\Sigma}
 \left( R^{\, t}_{\ t} +\Omega_1 R^{\, t}_{\ \vphi} +\Omega_2 R^{\, t}_{\ \psi}
 \right)
 \sqdetg \ dr d\theta d\vphi d\psi 
\ . \end{eqnarray}
We then express the Ricci tensor in the integrand of $I$
in terms of the stress-energy tensor,
\begin{eqnarray}
I
&=&  - \frac{2}{3}\int_{\Sigma} \left[
 (F_{t \alpha}+\Omega_1 F_{\vphi \alpha} +\Omega_2 F_{\psi \alpha})F^{t \alpha}
\right]
\sqdetg dr d\theta d\vphi d\psi 
\phantom{dr d\theta d\vphi d\psi dr d\psi}
\nonumber \\
 &+&  \frac{1}{3}\int_{\Sigma} \left[
 F_{\vphi\alpha}(F^{\vphi \alpha}-\Omega_1 F^{t \alpha})
+F_{\psi\alpha}(F^{\psi \alpha}-\Omega_2 F^{t \alpha})
\right]
\sqdetg dr d\theta d\vphi d\psi
\ . \end{eqnarray}
To evaluate the integral $I$ we make use of the Maxwell equations and
employ $F_{t \alpha}=-\partial_{\alpha} A_t$,
$F_{\vphi \alpha}=-\partial_{\alpha} A_{\vphi}$, and
$F_{\psi \alpha}=-\partial_{\alpha} A_{\psi}$, thus
\begin{eqnarray}
I
&=& \frac{2}{3}\int_{\Sigma} \partial_\alpha \left[
 (A_{t}+\Omega_1 A_{\vphi} + \Omega_2 A_{\psi} )F^{t \alpha}
\sqdetg \right] dr d\theta d\vphi d\psi
\phantom{dr d\theta d\vphi d\psi dr d\theta d\vphi d\psi}
\nonumber \\
&-& \frac{1}{3}\int_{\Sigma} \partial_\alpha \left[
\left( A_{\vphi}(F^{\vphi \alpha}-\Omega_1 F^{t \alpha})
     + A_{\psi}(F^{\psi \alpha}-\Omega_2 F^{t \alpha}) \right)
\sqdetg \right] dr d\theta d\vphi d\psi
\ . \end{eqnarray}
Exploiting the expansions for the metric and the gauge potential,
we note, that
the only non-vanishing contribution to the integral $I$ comes from
its first part, evaluated at the horizon,
\begin{equation}
I = - \frac{2}{3} (2 \pi)^2 \int_{0}^{\pi /2} d\theta 
\left. \left( \sqdetg (A_t +\Omega_1 A_{\vphi} +\Omega_2 A_{\psi}) 
 F^{t r} \right) \right|_{r_{\rm H}}
\ . \end{equation}
Since the electrostatic potential $\Phi_{\rm H}$,
Eq.~(\ref{esp}),
is constant at the horizon, 
\begin{equation}
I = - \frac{2}{3} (2 \pi)^2 \Phi_{\rm H} \int_{0}^{\pi /2} d\theta
\left. \left( \sqdetg F^{t r} \right) \right|_{r_{\rm H}} = 
\frac{2}{3} \Phi_{\rm H} Q
\ , \end{equation}
where $Q$ is the electric charge \cite{string3}. 
This yields the mass formula (\ref{mass})
$$
\frac{2}{3}M=
\frac{\kappa A_{\rm H}}{8\pi G}+\Omega_1 J_1 +\Omega_2 J_2
  + \frac{2}{3} \Phi_{\rm H} Q
\ . 
$$

\section{Numerical results}

To obtain asymptotically flat solutions, we impose
on the metric functions at infinity the boundary conditions
\begin{equation}
f=m=l=n=1 \ , \ \ \ p=\omega_1=\omega_2=0
\ . \label{bc1a} \end{equation}
For the gauge potential we choose a gauge where it vanishes at infinity,
\begin{equation}
A_t=A_\vphi=A_\psi=0
\ . \label{bc1b} \end{equation}

The horizon is located at $r_{\rm H}$,
and is characterized by the condition $f(r_{\rm H})=0$ \cite{kkrot,kkrot_c}.
Requiring the horizon to be regular, we obtain for the metric functions
the boundary conditions
\begin{equation}
f=m=l=n=p=0 \ , \ \ \ \omega_1=\omega_{1,\rm H}
\ , \ \ \ \omega_2=\omega_{2,\rm H}
\ , \label{bh2a} \end{equation}
where $\omega_{i,\rm H}$ are constants
determining the horizon angular velocities
$\Omega_i=\omega_{i,\rm H}/r_{\rm H}$.
The gauge potential satisfies
\begin{equation}
\left. \chi^\mu A_\mu \right|_{r_{\rm H}} = 
\Phi_{\rm H} \ , \ \ \ 
\partial_r A_\vphi=0 \ , \ \ \
\partial_r A_\psi=0
\ . \label{bh2b} \end{equation}

The boundary conditions in the planes $\theta = 0$ and $\theta = \pi /2$
are determined by symmetries. The metric functions satisfy
\begin{equation}
\partial_{\theta}f=\partial_{\theta}m=
\partial_{\theta}l=\partial_{\theta}n=
\partial_{\theta}p=
\partial_{\theta}\omega_1=
\partial_{\theta}\omega_2=0
\ , \label{bh3a} \end{equation}
while the gauge field functions satisfy at $\theta = 0$
\begin{equation}
\partial_\theta A_t=\partial_\theta A_\psi=0 \ , \ \ \ A_\vphi=0
\ , \label{bh3b} \end{equation}
and at $\theta = \pi /2$
\begin{equation}
\partial_\theta A_t=\partial_\theta A_\vphi=0 \ , \ \ \ A_\psi=0
\ . \label{bh3c} \end{equation}

For the numerical calculations we introduce
the compactified radial variable $\bar{r}= 1-r_{\rm H}/r$ \cite{kkrot,kkrot_c}.
We perform the numerical calculations with help of the package FIDISOL 
\cite{fidi}, based on the Newton-Raphson method.
For black holes with two independent angular momenta, we generically have 
to solve a system of 10 partial differential equations.
In the special case of black holes with a single angular momentum,
we retain only a system of 7 partial differential equations,
since the functions $p$, $\omega_2$, and $A_\psi$ vanish identically.
In the second special case of black holes with two equal-magnitude angular
momenta the angular dependence of the functions can be extracted
\cite{string2},
leaving a system of 6 ordinary differential equations \cite{long}. 
These black holes are therefore obtained with much higher
numerical accuracy.

Turning to the numerical results,
we focus here on black holes with a single angular momentum
and black holes with two equal-magnitude angular momenta.
In both cases we need to consider only a single function
$\omega(r)=\omega_1(r)$.
As we increase $\Omega$ from zero, while keeping $r_{\rm H}$ 
and $Q$ fixed,
a branch of rotating black hole solutions emerges from
the corresponding static black hole.
This lower branch extends up to a maximal value of $\Omega$,
which depends on $r_{\rm H}$ and $Q$.
From the maximal value $\Omega_{\rm max}$
a second branch, the upper branch, bends backwards towards
$\Omega=0$.
Along both branches mass and angular momentum 
continuously increase, as seen in Fig.~1,
where we exhibit the mass $M$ and the angular momentum $J=J_1$
for black holes with a single angular momentum
and black holes with two equal-magnitude angular momenta
as functions of $\Omega$ for $r_{\rm H}=0.5$,  
and $Q=\alpha \pi^{3/2}$, $\alpha=0$, 1, 3 and 5.
\begin{figure}[h!]
\parbox{\textwidth}
{\centerline{
\mbox{
\epsfysize=10.0cm
\includegraphics[width=70mm,angle=0,keepaspectratio]{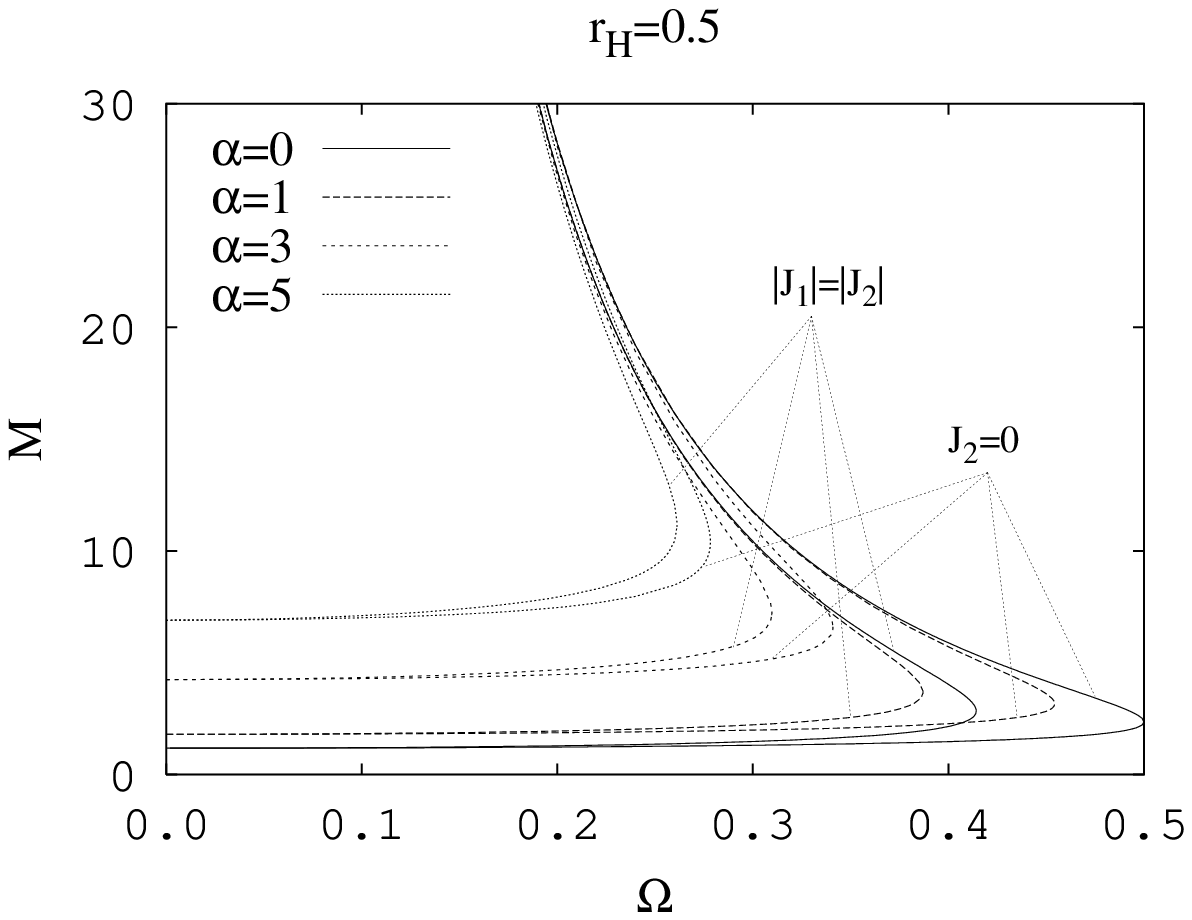}
\includegraphics[width=70mm,angle=0,keepaspectratio]{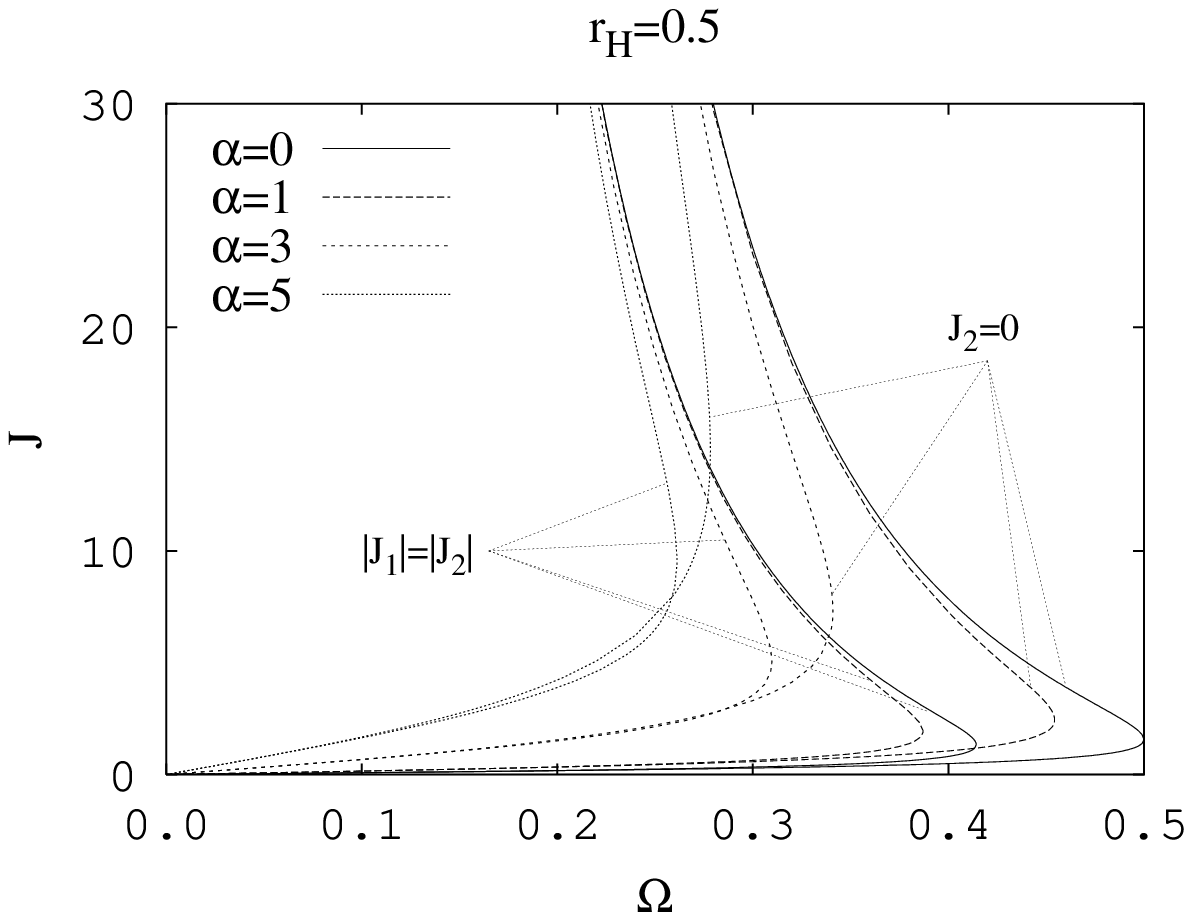}
}}}
\caption{
The mass $M$ and the angular momentum $J$
of black holes with a single angular momentum
and black holes with two equal-magnitude angular momenta
are shown as functions of the horizon angular velocity $\Omega$
for the horizon radius $r_{\rm H}=0.5$ 
and electric charge $Q=\alpha \pi^{3/2}$, $\alpha=1$, 3 and 5.
For comparison the corresponding values of the Myers-Perry 
black holes are also shown ($\alpha=0$).
}
\end{figure}

In Fig.~2 we show the horizon area parameter $r_\Delta$
as well as the surface gravity $\kappa$ for
the same set of solutions.
The horizon area increases monotonically along both branches,
and the surface gravity decreases monotonically,
tending to zero for large black holes.
\begin{figure}[h!]
\parbox{\textwidth}
{\centerline{
\mbox{
\epsfysize=10.0cm
\includegraphics[width=70mm,angle=0,keepaspectratio]{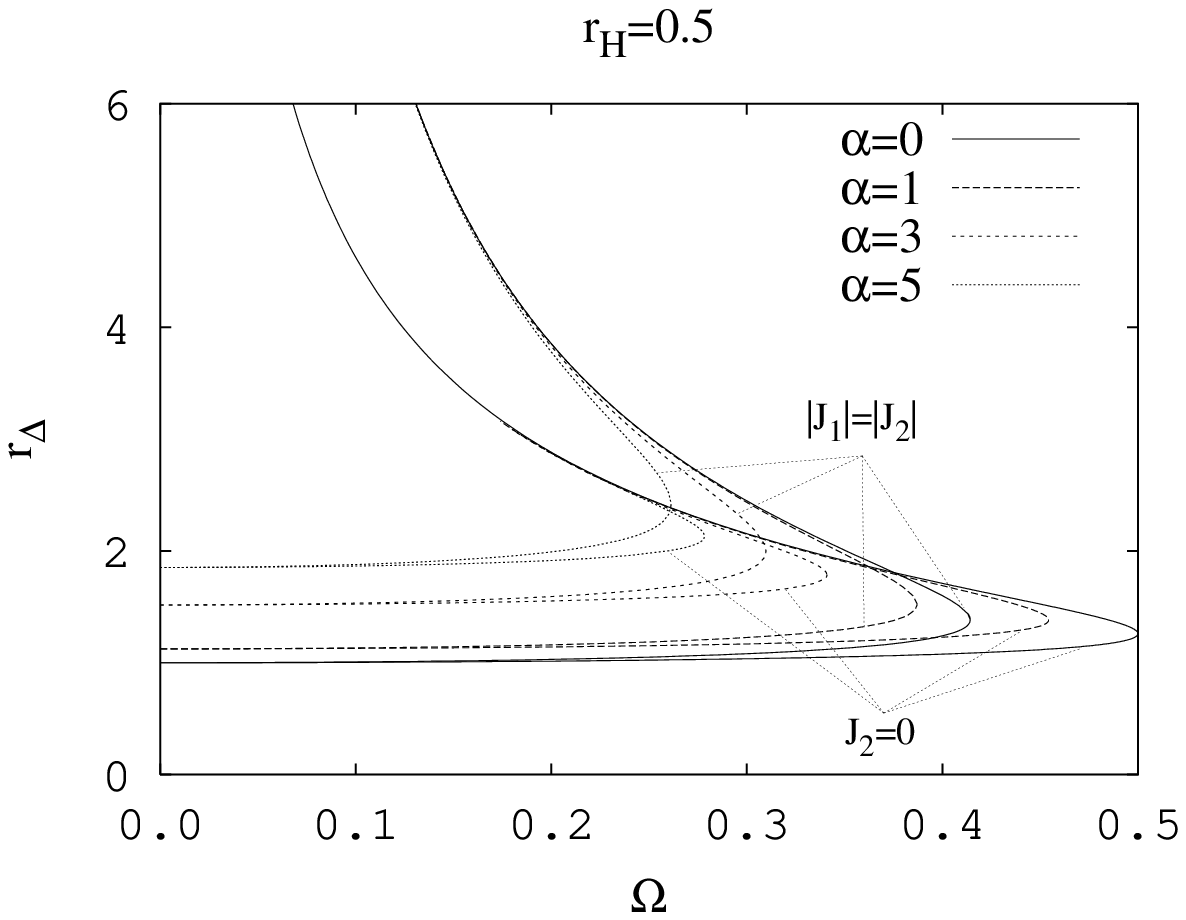}
\includegraphics[width=70mm,angle=0,keepaspectratio]{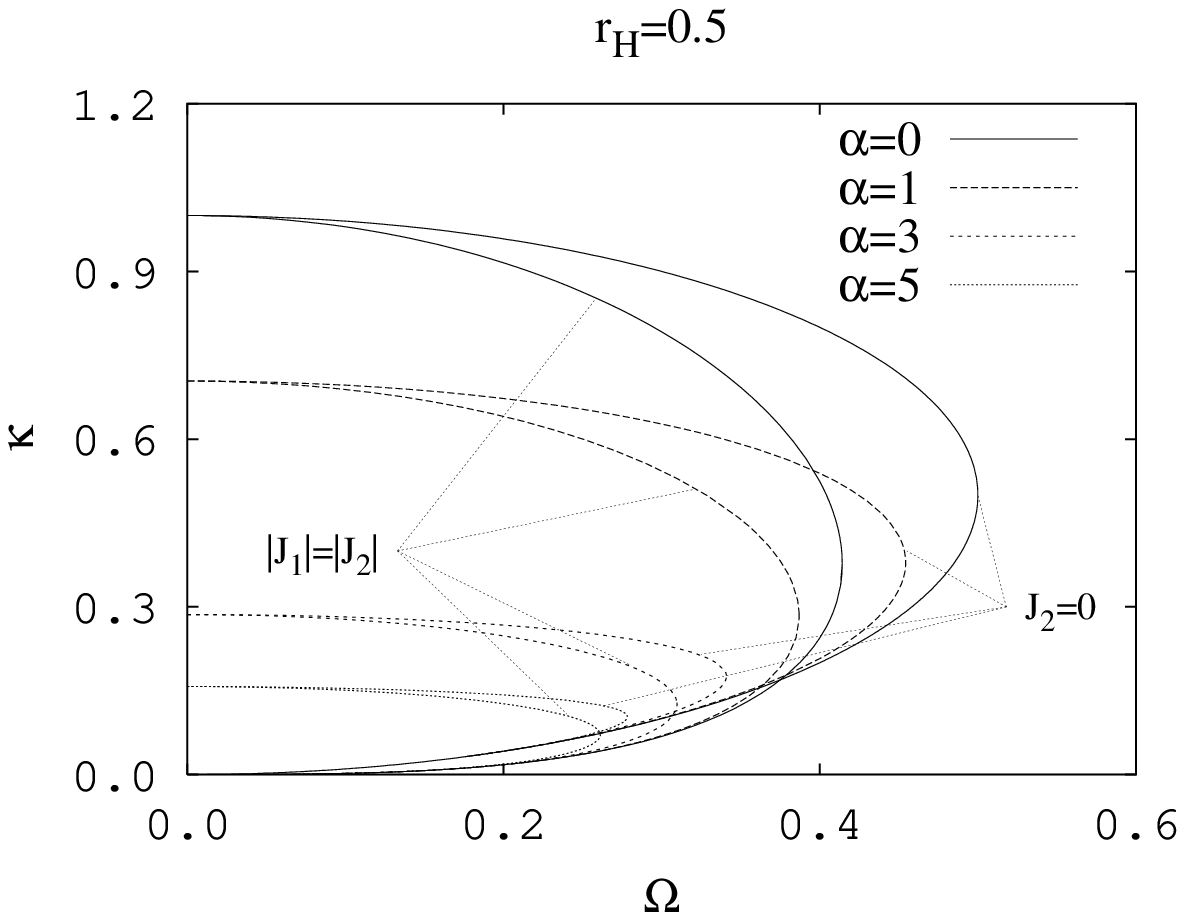}
}}}
\caption{
Same as Fig.~1 for the area parameter $r_\Delta$
and the surface gravity $\kappa$.
}
\end{figure}

In Fig.~3 we show the horizon electrostatic potential $\Phi_{\rm H}$
and the gyromagnetic ratio $g=g_1$
for these solutions.
While close to the perturbative value $g=3$ \cite{frolov},
we observe deviations from this value
of up to a few percent
for larger values of $Q$. Interestingly,
for black holes with a single angular momentum
these deviations are negative, while they are positive
for black holes with two equal-magnitude angular momenta \cite{foot}.
\begin{figure}[h!]
\parbox{\textwidth}
{\centerline{
\mbox{
\epsfysize=10.0cm
\includegraphics[width=70mm,angle=0,keepaspectratio]{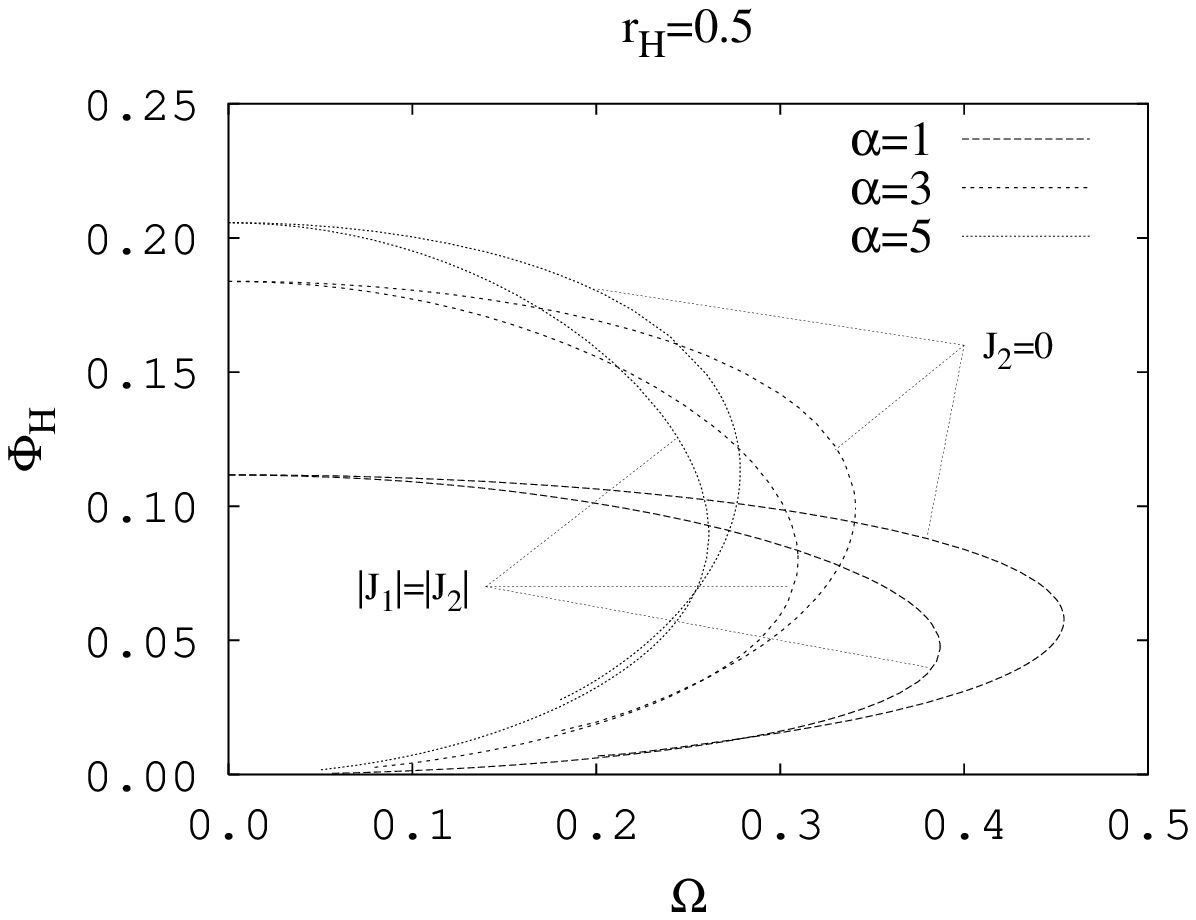}
\includegraphics[width=70mm,angle=0,keepaspectratio]{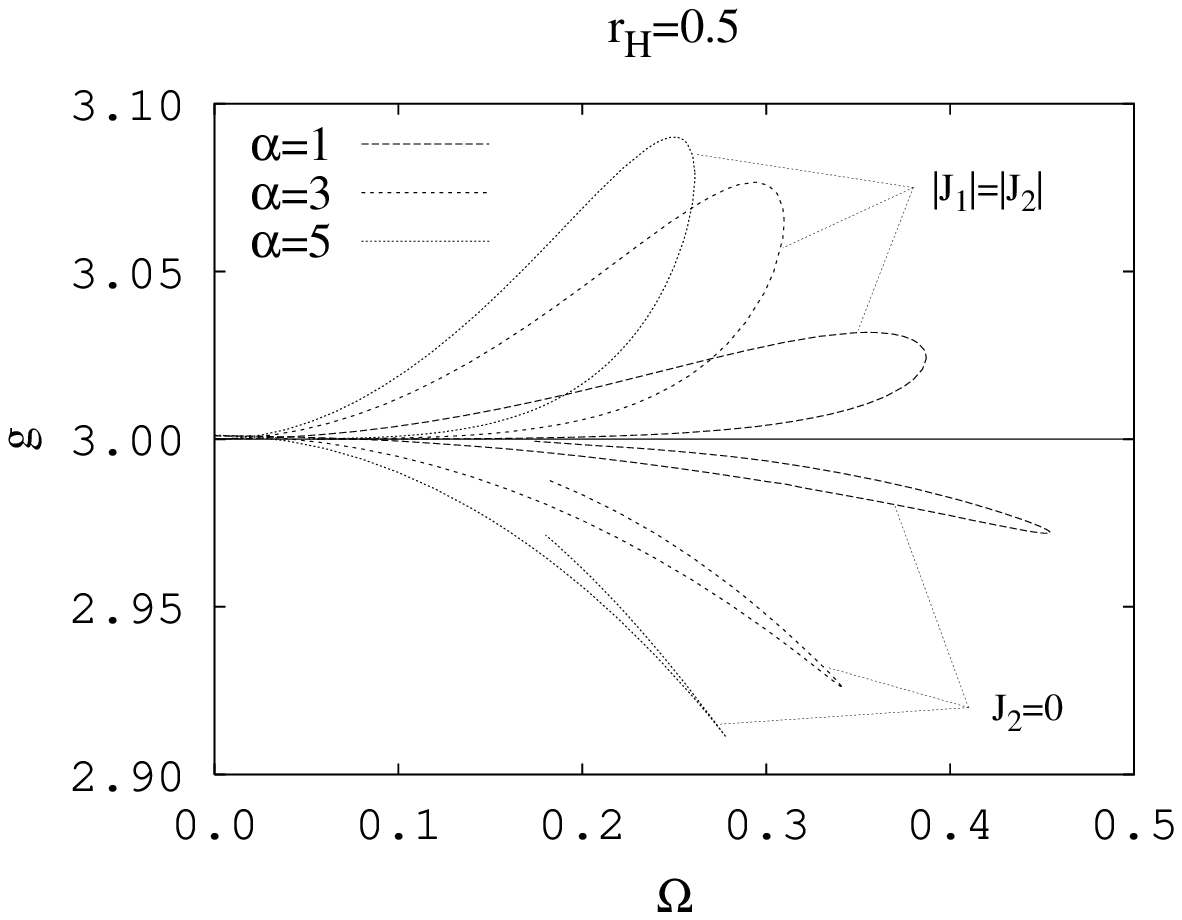}
}}}
\caption{
Same as Fig.~1 for the horizon electrostatic potential $\Phi_{\rm H}$ 
and the gyromagnetic ratio $g$.
}
\end{figure}
 
Finally, in Fig.~4 we show the mass $M$ 
and the angular momentum $J$
for black holes with a single angular momentum
and black holes with two equal-magnitude angular momenta
as functions of the isotropic horizon radius $r_{\rm H}$
for $\omega_{\rm H}=0.1$ and 
$Q=\alpha \pi^{3/2}$, $\alpha=0$, 1, 3 and 5.
\begin{figure}[h!]
\parbox{\textwidth}
{\centerline{
\mbox{
\epsfysize=10.0cm
\includegraphics[width=70mm,angle=0,keepaspectratio]{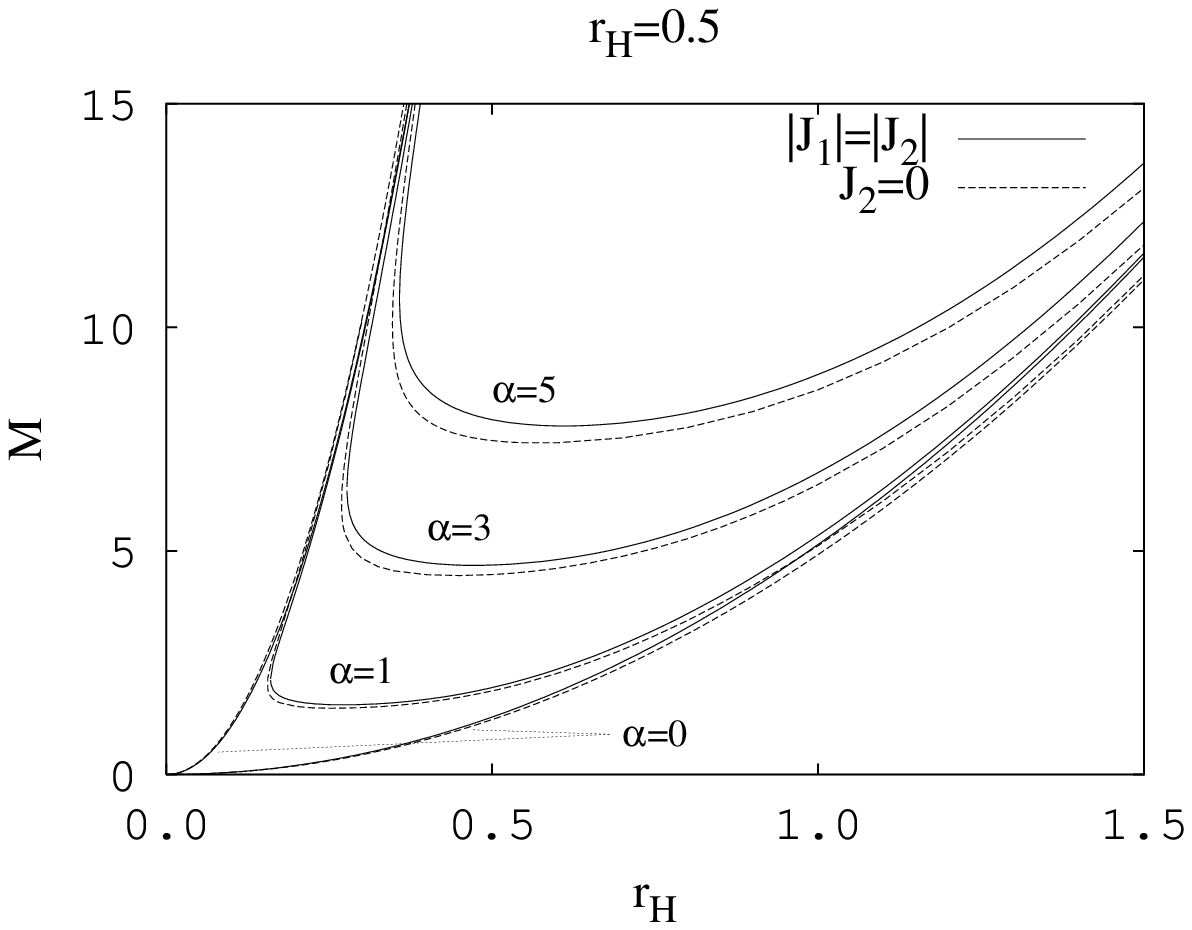}
\includegraphics[width=70mm,angle=0,keepaspectratio]{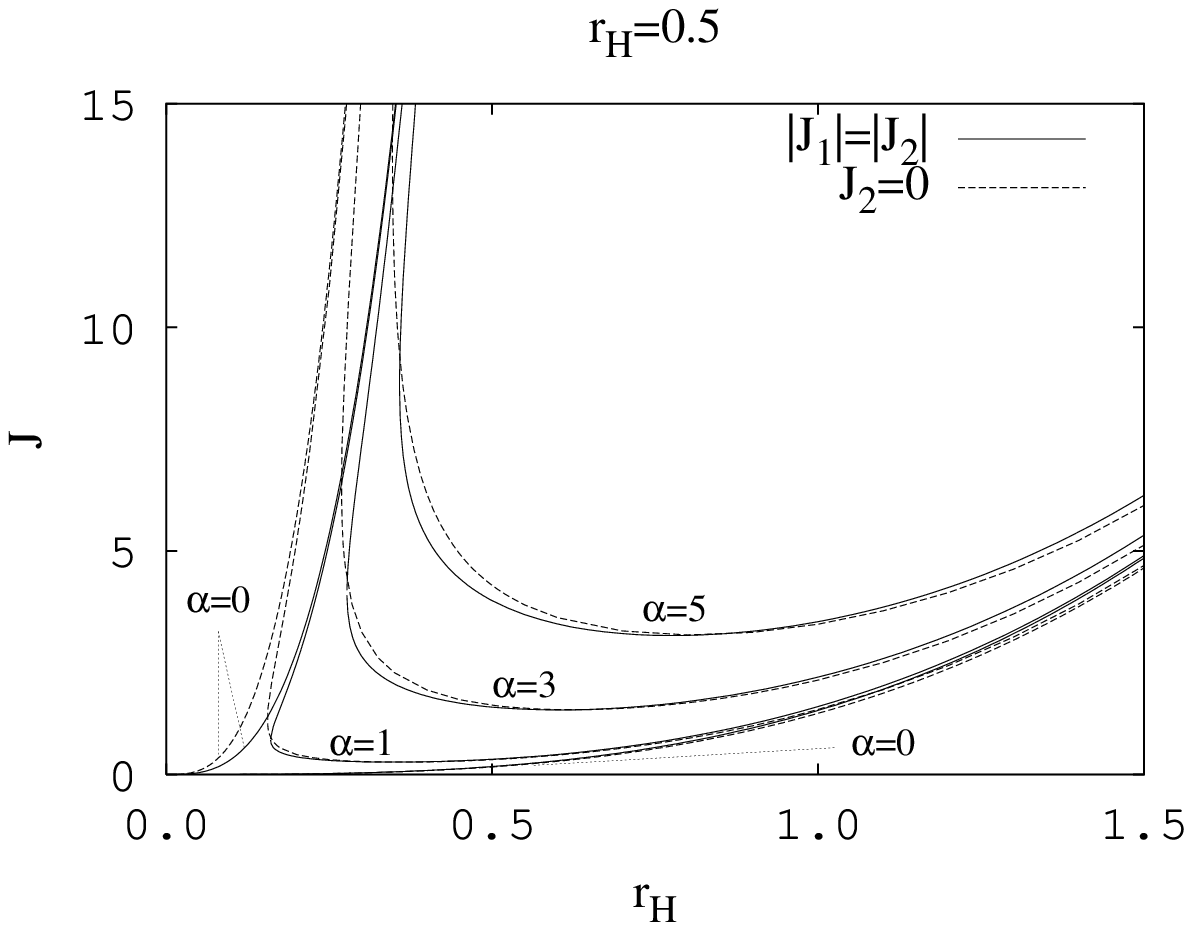}
}}}
\caption{
The mass $M$ and the angular momentum $J$
of black holes with a single angular momentum
and black holes with two equal-magnitude angular momenta
are shown as functions of the horizon radius $r_{\rm H}$
for $\omega_{\rm H}=0.1$ 
and electric charge $Q=\alpha \pi^{3/2}$, $\alpha=1$, 3 and 5.
For comparison the corresponding values of the Myers-Perry 
black holes are also shown ($\alpha=0$).
}
\end{figure}

Further details and numerical results for the general case of
two nonequal-magnitude angular momenta will be given elsewhere \cite{long}.

\section{Conclusions}

We have considered charged rotating black holes in 
5-dimensional Einstein-Maxwell theory.
These black holes are asymptotically flat,
they possess a regular horizon of spherical topology
and two independent angular momenta
associated with two distinct planes of rotation.
Our appoach has been analytical concerning the expansions
at infinity, at the horizon, and in the symmetry planes,
and numerical concerning the actual construction of the black holes.
In particular, we have introduced a metric parametrization, 
based on bi-azimuthal isotropic coordinates,
well suited for numerical work.

We have studied the physical properties of these black holes,
in particular their global charges and horizon properties,
and obtained a generalized Smarr formula for these black holes.
Interestingly, for generic values of the charge
their gyromagnetic ratios differ from the value $g_i=3$, 
obtained for weakly charged EM black holes \cite{frolov}
(or certain supersymmetric black holes \cite{herdeiro}).

In the numerical construction of the black holes
we here have focussed on black holes with a single angular momentum 
and black holes with two equal-magnitude angular momenta.
In the latter case, the angular dependence of the metric and
gauge field functions is obtained explicitly, leaving only
a set of ordinary differential equations to be solved numerically.
Further details of these black hole solutions
and numerical results for black holes with
two nonequal-magnitude angular momenta
will be presented elsewhere \cite{long}.

While analytical construction of the exact
5-dimensional charged rotating EM black holes
still presents an outstanding challenge, 
we hope, that the numerical solutions presented here
will be helpful in achieving this goal.

The Ansatz introduced to parametrize the metric
of the 5-dimensional rotating EM black holes
should also allow for the numerical construction of
rotating black holes of theories containing further matter fields.
In particular, the inclusion of a dilaton might lead to 
interesting phenomena. 
Already in $3+1$ dimensions the presence of a dilaton yields 
new surprising features,
such as counterrotating black holes \cite{kkrot_c}.

{\sl Acknowledgement}

FNL gratefully acknowledges Universidad Complutense de Madrid for
support under Proyecto Complutense PR1/05-13355.


\begin{thebibliography}{000}
\bibitem{foot}
We contribute the very slight excess of the gyromagnetic ratio 
over the perturbative value $g=3$, seen for
black holes with a single angular momentum, small charge and small
horizon angular velocity, to numerical inaccuracy, since
for these black holes
a system of partial differential equations has to be solved.
\end{thebibliography}
\end{document}